\documentclass[preprint,12pt]{elsarticle}

\usepackage{amssymb}

\usepackage{verbatim}
\usepackage{longtable}
\usepackage{xspace}

\newcommand*{\wn}{cm$^{-1}$\xspace}
\newcommand*{\um}{$\mu$m\xspace}
\newcommand*{\etal}{\emph{et al.}\xspace}
\newcommand*{\abin}{\emph{ab initio}\xspace}
\newcommand*{\xstate}{$X\,^{1}\Sigma_{g}^{+}$\xspace}
\newcommand*{\efstate}{$EF\,^{1}\Sigma^{+}_{g}$\xspace}
\newcommand*{\gkstate}{$GK\,^{1}\Sigma^{+}_{g}$\xspace}
\newcommand*{\bstate}{$B\,^1\Sigma^{+}_{u}$\xspace}
\newcommand*{\bpstate}{$B'\,^1\Sigma^{+}_{u}$\xspace}
\newcommand*{\Hm}{H$_{2}$\xspace}
\newcommand*{\Dm}{D$_{2}$\xspace}
\newcommand*{\EFX}{$EF\,{}^{1}\Sigma_{g}^{+}-X\,{}^{1}\Sigma_{g}^{+}$\xspace}

\newcommand*{\BX}{$B\,{}^{1}\Sigma_{u}^{+}-X\,{}^{1}\Sigma_{g}^{+}$\xspace}
\def\cjp{Can.\ J. Phys.\ }

\def\jcp{J. Chem.\ Phys.\ }
\def\jctc{J.\ Chem.\ Theory\ Comput.\ }

\def\jms{J. Mol.\ Spectrosc.\ }

\def\jpcrd{J. Phys. Chem. Ref. Data }

\def\jqsrt{J. Quant.\ Spectr.\ Radiat.\ Transfer }
\def\molp{Mol.\ Phys.\ }

\def\pra{Phys.\ Rev.\ A }

\def\prl{Phys.\ Rev.\ Lett.\ }

\journal{Journal of Molecular Spectroscopy}

\begin{document}

\begin{frontmatter}



\title{$B\,^1\Sigma^{+}_{u}$ and $EF\,^{1}\Sigma^{+}_{g}$ level energies of D$_2$}

\author[VU,USC]{E. J. Salumbides}
\author[Orsay]{D. Bailly}
\author[Soleil]{M. Vervloet}
\author[VU]{W. Ubachs}

\address[VU]{Department of Physics and Astronomy, LaserLaB, VU University,\\ de Boelelaan 1081, 1081HV Amsterdam, the Netherlands}
\address[USC]{Department of Physics, University of San Carlos, Cebu City 6000, Philippines}
\address[Orsay]{Laboratoire Photophysique Mol\'{e}culaire, Universit\'{e} de Paris-Sud, Orsay, France}
\address[Soleil]{Synchrotron Soleil, L’orme des Merisiers, Saint-Aubin BP 48,\\ 91192 Gif-sur-Yvette, France}

\begin{abstract}

Accurate absolute level energies of the $B\,^1\Sigma^{+}_{u}$, $v=0-8, N$ and $EF\,^{1}\Sigma^{+}_{g}$, $v=0-21, N$ rovibrational quantum states of molecular deuterium are derived by combining results from a Doppler-free two-photon laser excitation study on several lines in the $EF\,{}^{1}\Sigma_{g}^{+}-X\,{}^{1}\Sigma_{g}^{+}$ (0,0) band, with results from a Fourier-transform spectroscopic emission study on a low-pressure hydrogen discharge.
Level energy uncertainties as low as 0.000\,5 cm$^{-1}$ are obtained for some low-lying  $E\,^{1}\Sigma^{+}_{g}$ inner-well rovibrational levels, while uncertainties for higher-lying rovibrational levels and those of the $F\,^{1}\Sigma^{+}_{g}$ outer-well states are nominally 0.005 cm$^{-1}$.
Level energies of $B\,^1\Sigma^{+}_{u}$ rovibrational levels, for $v \leq 8$ and $N \leq 10$ are determined at an accuracy of 0.001 \wn.
Computed wavelengths of D$_2$ Lyman transitions in the $B\,^1\Sigma^{+}_{u}-X\,^{1}\Sigma^{+}_{g}$ ($v,0$) bands are also tabulated for future applications.

\end{abstract}


\end{frontmatter}


\section{Introduction}
\label{intro}

Molecular hydrogen continues to be of relevance as it is the simplest neutral system yielding the most accurate results from molecular quantum theory.
This benchmark molecule offers a natural setting for the confrontation of the most advanced first-principles theoretical calculations with accurate experimental investigations.
Stable isotopologues of molecular hydrogen, \Hm, HD, and \Dm are treated in analogous calculations, except for the additional g/u symmetry breaking in HD \cite{DeLange2002}.
For example, sustained efforts throughout the decades, in both the theoretical and experimental realm, have led to remarkable improvements in the determination of the dissociation energy $D_0$ of the molecular hydrogen isotopologues \Hm~\cite{Liu2009}, HD~\cite{Sprecher2010}, and \Dm~\cite{Liu2010}.
At the present level of accuracy, subtle QED effects in molecular level energies need to be accounted for in calculations to obtain agreement with the measurements. Such calculations have now been performed by Pachucki and coworkers for the \xstate electronic ground state of all three isotopomers yielding $10^{-3}$ \wn\ uncertainties in the binding energies for bound rovibrational quantum states~\cite{Piszcziatowski2009,Komasa2011}. However, for excited states as the \bstate and the \efstate such accurate calculations have not been performed yet.

While \Hm and HD are of great interest in astronomical and cosmological investigations, see e.g. \cite{Weerdenburg2011}, \Dm has not been observed in space beyond the solar system.
Since deuterium and tritium will be the main fuel for experimental fusion reactors, their spectra are important diagnostic tools to study the various molecular hydrogen isotopologues produced in the nuclear reactions.
These \Dm and T$_2$ fuels will be heated to extremely high temperatures, so that it is important to characterize the plasma dynamics, for which the excitation cross-sections induced either by photons or electrons are of relevance~\cite{Liu2012b}.

The present work focuses on the determination of accurate level energies for the \bstate and the \efstate states in the D$_2$ isotopologue. It builds on the long tradition of studies of the Lyman bands, associated with the \BX system, that includes the strongest transitions in the molecule. Early, classical spectroscopic studies of the Lyman bands of D$_2$ were performed by Herzberg and coworkers~\cite{Bredohl1973,Dabrowski1974}, later followed by XUV-laser spectroscopic studies at increasing resolution and precision~\cite{Hinnen1994,Roudjane2008}, albeit only for a relatively small subset of bands in the Lyman system.
More comprehensive studies, delivering spectroscopic information on a large set of rovibrational levels, involved electron scattering induced emission studies~\cite{Abgrall1999} (with a full database provided~\cite{D2trans}), laser probing of highly excited rovibrational levels in a plasma~\cite{Gabriel2009}, and a VUV Fourier-transform absorption study~\cite{DeLange2012}.

The \efstate state, which is long-lived, since one-photon decay to the ground state is dipole forbidden, has been subject of many investigations over the years.  Freund~\etal\cite{Freund1985} compiled a comprehensive data set of D$_2$ level energies from the studies performed by Dieke over many decades. Yu and Dressler have assembled the information, derived from classical spectroscopy, on the D$_2$ \efstate state in a comparison to an \abin model~\cite{Yu1994}. Two-photon laser excitation of the \efstate double-well state was vigorously pursued leading to increased precision of level energies~\cite{Kligler1978,Yiannopoulou2006,Hannemann2006}. The most recent studies targeted excitation from the \xstate, $v=1$ level for a sensitive test of QED calculations in the D$_2$ ground electronic state~\cite{Dickenson2013,Niu2014}.
The analogous 2+1 resonance-enhanced multi-photon ionization study by Heck~\etal was performed at much lower resolution and accuracy but provided information on a larger manifold of rovibrationally excited states~\cite{Heck1995}.

In the present investigation, accurate level energies of the \bstate and \efstate electronic states for \Dm are derived from new high-resolution Fourier-transform (FT) spectroscopic data. These level energies are anchored with respect to the ground state \xstate using accurate $EF-X$ transitions from previous UV spectroscopy investigations~\cite{Hannemann2006,Dickenson2013,Niu2014}.
The FT spectroscopy setup has been discussed in detail in Ref.~\cite{Bailly2007}, while the anchoring method for the derivation of accurate level energies in the \efstate and \bstate electronically excited states has been discussed in previous studies focusing on H$_2$~\cite{Salumbides2008,Bailly2010}. Here, the calibration procedures on the FT-spectra have been improved mainly by referencing against updated wavelength standards~\cite{Whaling1995,Whaling2002,Saloman2010}.

This work results in level energies for a large set of quantum levels for the \bstate, $v=0-8$ and \efstate, $v=0-21$ vibrational states, being the most accurate to date. For future use, the transition wavelengths of the Lyman bands are calculated based on presently determined level energies and those of the \xstate ground electronic state~\cite{Piszcziatowski2009,Komasa2011}.

\section{Method}

\begin{figure}
\center
\resizebox{0.7\textwidth}{!}{\includegraphics{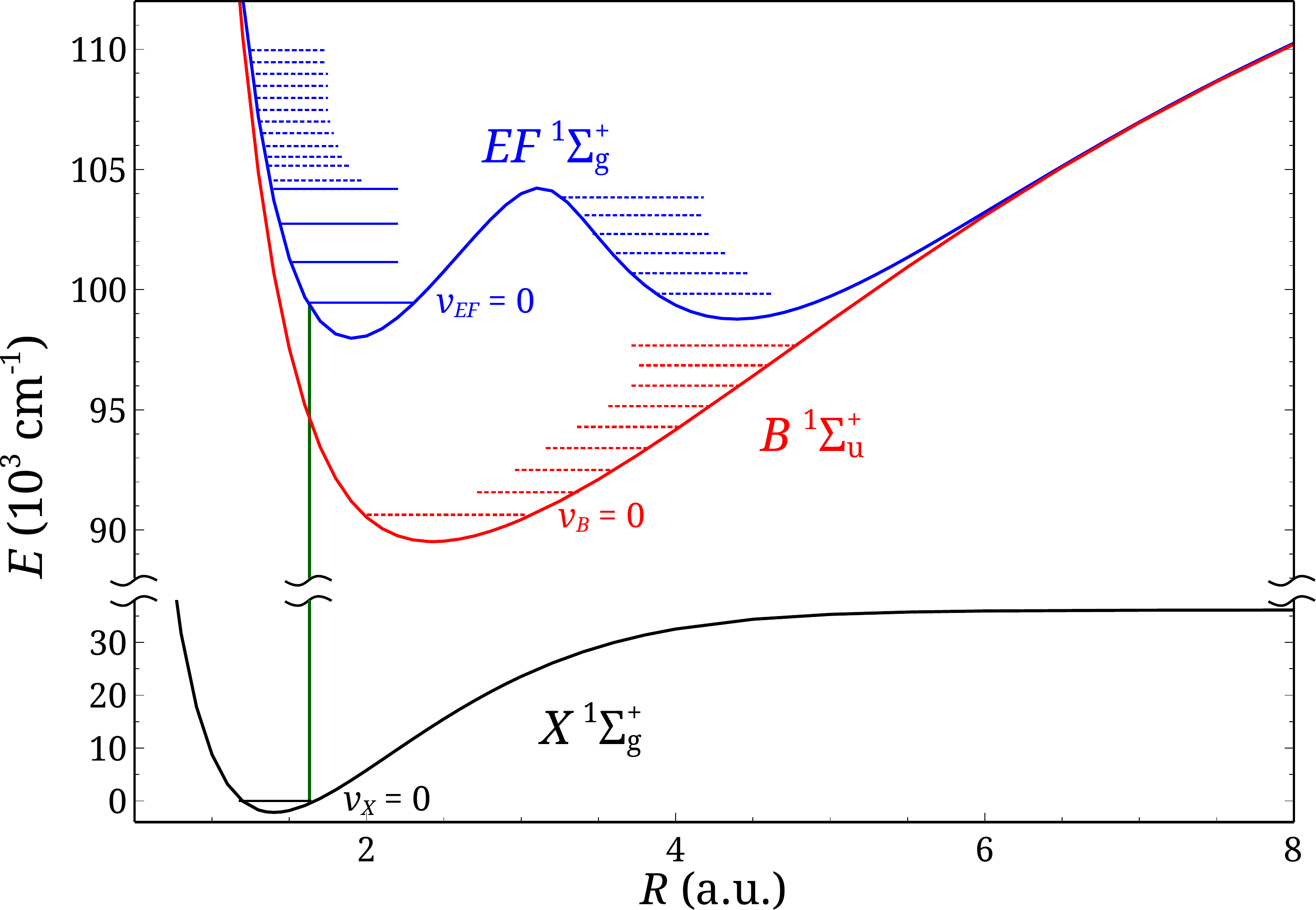}}
\caption{
Potential energy curves of \Dm\ electronic states relevant to this study. The anchor transition connecting the manifold of \efstate\ and \bstate\ level energies to the \xstate\ ground state is indicated by the vertical line.
The rotationless term values for the \efstate\ and \bstate\ vibrational states presently analyzed are also indicated.
}
\label{Potential}
\end{figure}

The experimental determination of \bstate and \efstate level energies is based on an approach that was previously employed for \Hm~\cite{Salumbides2008,Bailly2010}.
Accurate values are derived from two completely independent spectroscopic investigations. Relative level energies of a wide manifold of rovibrational quantum states in the \bstate and \efstate states are determined via high resolution FT emission spectroscopy of transitions connecting $B$, $EF$, \bpstate, and \gkstate electronically excited states of \Dm.
Levels from both the inner and outer wells of the \efstate state are covered (see Fig.~\ref{Potential}).

The relative energies are brought onto an absolute energy scale by anchoring them to the \xstate electronic ground state by results from two-photon UV laser spectroscopy on the \EFX system~\cite{Hannemann2006}. The general features of this measurement scheme include Doppler-free two-photon ionisation and use of a narrowband pulsed titanium-sapphire laser with pulse frequency chirp measurement and correction~\cite{Hannemann2007a}, as well as absolute frequency calibration against a frequency comb laser.
These two-photon UV laser experiments yield highly accurate level energies for two anchor levels used in the present \Dm study~\cite{Hannemann2006}:
the \EFX Q(0) transition energy of 99\,461.449\,08\,(1) \wn for ortho-\Dm, and the \EFX Q(1) transition energy at $99\,433.716\,38\,(10)$ \wn for para-\Dm.
The ortho-\Dm levels have total nuclear spin $I_T=0,2$, while para-\Dm have $I_T=1$. For the \efstate electronic state symmetry, the even-$N$ rotational levels belong to ortho-\Dm while the odd-$N$ levels to para-\Dm.
For the latter para-state anchor transition, we have used in addition the \Dm \xstate, $v=0, N=1$ level energy of $59.780\,615\,(3)$ \wn from the accurate theoretical calculations of Komasa~\etal~\cite{Komasa2011}.
The entire manifold of \Dm excited states becomes anchored to the \xstate $v=0, N=0$ ground state with the use of the ortho and para anchor levels.
The same two-photon UV experimental scheme has recently been employed for an accurate determination of the \efstate anchor against the $v=1$ level in the \xstate ground state~\cite{Dickenson2013,Niu2014}, thus verifying the values for the anchor levels.

\begin{figure}
\center
\resizebox{0.7\textwidth}{!}{\includegraphics{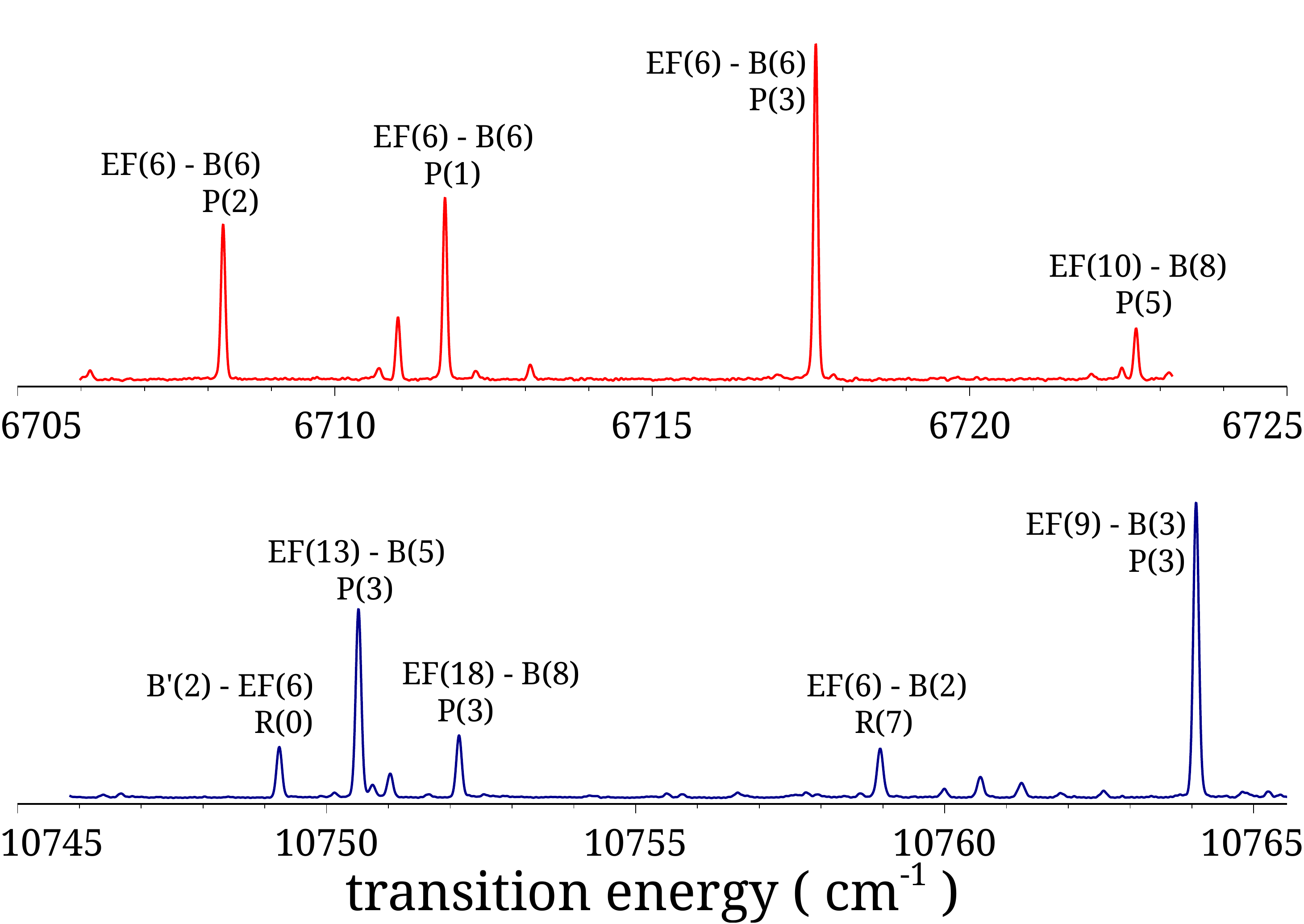}}
\caption{
Fourier-transform spectra showing transitions connecting the \Dm\ \efstate\ and \bstate\ electronic states. The two ranges represent small slices of the whole FT spectral range spanning from $1\,800$ to $20\,000$ \wn.
}
\label{FT_spectra}
\end{figure}

In the FT emission investigation, an extremely broad wavelength range is covered, from 450 nm in the visible to 5 \um in the infrared. The entire spectral data set includes a multitude of transitions belonging to mutually overlapping band systems.
The experimental details have been described previously~\cite{Bailly2007,Bailly2010} and only the general features are recalled here.
A low-pressure microwave discharge ($f\sim2450$ MHz) was established in a quartz tube (diameter: 1 cm, length: 25 cm) where molecular deuterium flows through at moderate speed.
The microwave power (about 70 W) and the gas pressure (about 5 mbar) were controlled in order to optimize the optical emission intensity of the molecular species (relative to the atomic emissions), as well as to maintain stability of the fluorescence.
The discharge emission was focused onto the entrance iris of a Bruker IFS 120 FT spectrometer.
The emission spectrum was recorded from $1\,800$ to $22\,000$ \wn, with the data collection subdivided into smaller spectral range recordings using appropriate coloured or optical interference filters and detectors.
Sample FT spectra are displayed in Fig.~\ref{FT_spectra} showing the fluorescence transitions connecting some \Dm \efstate and \bstate levels.
The spectral resolution is limited by Doppler widths, varying from about 0.02 \wn (infrared) to 0.2 \wn (violet).

Traces of CO or Ar were introduced with the deuterium flow for wavelength calibration purposes.
The CO calibration transitions in the range between 1993 to 2254 \wn were referenced against the accurate microwave studies of Maki \etal \cite{Maki1990}, which have uncertainties of $2-5\times 10^{-5}$ \wn.
The Ar lines were referenced against the more recent results of Whaling \etal~\cite{Whaling1995,Whaling2002} with estimated uncertaintes of a few times $10^{-4}$ \wn.
We note that the Whaling values \cite{Whaling1995} for Ar II show deviations with the older Norlen database~\cite{Norlen1973} that can be as much as 0.01 \wn, specifically in the range from $17\,000$ to $21\,000$ \wn.
For the Ar I calibrations using \cite{Whaling2002}, the wavelengths must be corrected by a known factor as discussed by Sansonetti \cite{Sansonetti2007}.
A recent compilation by Saloman \cite{Saloman2010} for (neutral and ionic) Ar includes the most recent determinations.

The line assignment and rotational analyis of the FT spectra was performed with the aid of Ref.~\cite{Freund1985} that was based on Dieke's work.
The analysis was verified through the combination differences of the transition energies from different vibrational bands for each level.
For the \efstate level energies, transitions connecting $EF$ to the $B$ and \bpstate electronic states were used, while for the \bstate level energies, transitions from $B$ to $EF$ and \gkstate levels were used.
The analysis then yielded a consistent set of $EF$ and $B$ level energies with respect to the \efstate $v=0, N=0, 1$ levels.
Finally, the whole \efstate and \bstate level manifold is anchored to the \xstate $(v=0, N=0)$ ground state using the accurate \EFX (0,0) band transition energies from the VUV study of Hannemann \etal~\cite{Hannemann2006}.
Since several transitions from different vibrational bands were used in the derivation of each level energy, the accuracy depends on the number of transitions, as well as the signal-to-noise ratio, connected to that particular level.

\section{\efstate\ level energies}

The level energies of the \efstate\ state of \Dm\ for vibrational levels $v=0-21$ derived from FT spectroscopy are listed in Table~\ref{EF_levels}.
The ortho-\Dm (even $N$ levels) were anchored to the \xstate (0,0) level using the \efstate $v=0, N=0$ level energy of $99\,461.449\,08\,(10)$ \wn, while for para-\Dm (odd $N$ levels) the \efstate $v=0, N=1$ level energy of $99\,493.497\,00\,(10)$ \wn is used.
The $N=0-2$ rotational levels of $EF, v=0$ listed in Table~\ref{EF_levels} are derived from the results of Ref.~\cite{Hannemann2006} and ground state level energies of Komasa~\etal~\cite{Komasa2011}, although the $EF, v=0, N=2$ value was not used as an anchor level energy.
Note that the vibrational levels below the double-well barrier are labeled as belonging either to the inner $E$ or outer $F$ wells, in addition to the generalized $EF$ state vibrational quantum numbers.
The energies for levels involved in transitions with good signal-to-noise ratio are most accurate, e.g. vibrational bands belonging to the inner well labeled $E0, E1, E2$ and $E3$ with uncertainties as low as 0.000\,5 \wn with the least accurate at 0.005 \wn. The transitions belonging to levels within the outer $F$-well and the higher-lying vibrational bands are in general weaker and have larger uncertainties, mostly in the $0.001-0.006$ \wn range.

{\small
\renewcommand{\thefootnote}{\textdagger}
\begin{longtable}{cr@{.}l@{\hspace{20pt}}r@{.}l@{\hspace{20pt}}r@{.}l@{\hspace{20pt}}r@{.}l}
\caption{
Level energies (term values) in the $EF$ $^1\Sigma^+_g$ states of D$_2$ for vibrational levels $v = 0 -– 21$.
The vibrational assignment is given both in terms of a combined numbering in the $EF$ double-well potential (inside parentheses) and with alternative labels for vibrational bands bound within the separate $E-$ and $F-$wells.
The values are in cm$^{-1}$ with the $1\sigma$-uncertainties in between parenthesis ( ) expressed in units of the last digit.
\label{EF_levels}
}\\
\hline\hline\noalign{\smallskip}
$N$&\multicolumn{2}{c}{$(v=0)$ \emph{E0}}&\multicolumn{2}{c}{$(v=1)$ \emph{F0}}&\multicolumn{2}{c}{$(v=2)$ \emph{F1}}&\multicolumn{2}{c}{$(v=3)$ \emph{E1}}	\\
\hline\noalign{\smallskip}
\endfirsthead
\multicolumn{9}{c}
{\tablename\ \thetable\ -- \textit{Continued from previous page}} \\
\endhead
\hline \multicolumn{4}{l}{\footnotesize $^\dagger$ Derived from Ref.~\cite{Hannemann2006}} and Ref.~\cite{Komasa2011}  & \multicolumn{5}{r}{\textit{Continued on the next page}} \\
\endfoot
\hline\noalign{\smallskip}
\endlastfoot

0	& 99\,461&449\,1\,(1)$^\dagger$	&99\,829&136\,(5)		&100\,686&239\,(5)		&101\,149&667\,2\,(5)		\\
1	& 99\,493&497\,0\,(1)$^\dagger$	&99\,835&190\,(5)		&100\,692&225\,(5)		&101\,180&348\,8\,(5)		\\
2	& 99\,557&460\,6\,(1)$^\dagger$	&99\,847&581\,(5)		&100\,704&409\,(5)		&101\,241&580\,9\,(5)		\\
3	& 99\,653&077\,8\,(5)		&99\,865&990\,(5)		&100\,722&567\,(5)		&101\,333&102\,9\,(5)		\\
4	& 99\,779&959\,5\,(5)		&99\,890&577\,(5)		&100\,746&741\,(5)		&101\,454&528\,0\,(5)		\\
5	& 99\,937&593\,3\,(5)		&99\,921&241\,(5)		&100\,776&969\,(5)		&101\,605&196\,0\,(5)		\\
6	&100\,125&351\,0\,(5)		&\multicolumn{2}{c}{}		&100\,814&897\,(10)		&101\,784&934\,0\,(5)		\\
7	&100\,342&497\,0\,(5)		&\multicolumn{2}{c}{}		&\multicolumn{2}{c}{}		&101\,992&551\,5\,(5)		\\
8	&100\,588&195\,1\,(5)		&\multicolumn{2}{c}{}		&\multicolumn{2}{c}{}		&102\,227&362\,3\,(5)		\\
9	&100\,861&518\,0\,(5)		&\multicolumn{2}{c}{}		&\multicolumn{2}{c}{}		&102\,488&428\,1\,(5)		\\
10	&101\,161&453\,1\,(5)		&\multicolumn{2}{c}{}		&\multicolumn{2}{c}{}		&\multicolumn{2}{c}{}   	\\
11	&101\,486&939\,1\,(5)		&\multicolumn{2}{c}{}		&\multicolumn{2}{c}{}		&\multicolumn{2}{c}{}   	\\

\hline\noalign{\smallskip}
$N$&\multicolumn{2}{c}{$(v=4)$ \emph{F2}}&\multicolumn{2}{c}{$(v=5)$ \emph{F3}}&\multicolumn{2}{c}{$(v=6)$ \emph{E2}}&\multicolumn{2}{c}{$(v=7)$ \emph{F4}}	\\
\hline\noalign{\smallskip}									
0	&101\,516&074\,(5)		&102\,318&268\,(5)		&102\,741&656\,4\,(5)		&103\,091&713\,(2)	\\
1	&101\,522&042\,(5)		&102\,324&183\,(5)		&102\,770&786\,2\,(5)		&103\,097&588\,(2)    	\\
2	&101\,533&990\,(2)		&102\,336&000\,(5)		&102\,828&903\,6\,(5)		&103\,109&342\,(2)    	\\
3	&101\,551&890\,(2)		&102\,353&689\,(5)		&102\,915&721\,4\,(5)		&103\,126&972\,(2)   	\\
4	&101\,575&746\,(2)		&102\,377&272\,(5)		&103\,030&761\,5\,(5)		&103\,150&531\,(2)    	\\
5	&101\,605&675\,(2)		&102\,406&708\,(5)		&103\,170&003\,6\,(5)		&103\,183&462\,(2)    	\\
6	&\multicolumn{2}{c}{}		&\multicolumn{2}{c}{}		&103\,344&349\,4\,(5)		&103\,214&157\,(5)    	\\
7	&\multicolumn{2}{c}{}		&\multicolumn{2}{c}{}		&103\,540&466\,3\,(5)		&\multicolumn{2}{c}{}	\\
8	&\multicolumn{2}{c}{}		&\multicolumn{2}{c}{}		&103\,761&866\,5\,(20)		&\multicolumn{2}{c}{}	\\
9	&\multicolumn{2}{c}{}		&\multicolumn{2}{c}{}		&104\,006&094\,1\,(20)		&\multicolumn{2}{c}{}	\\
10      &\multicolumn{2}{c}{}           &\multicolumn{2}{c}{}           &104\,273&057\,1\,(50)          &\multicolumn{2}{c}{}	\\

\newpage

\hline\noalign{\smallskip}
$N$&\multicolumn{2}{c}{$(v=8)$ \emph{F5}}&\multicolumn{2}{c}{$(v=9)$ \emph{E3}}&\multicolumn{2}{c}{$(v=10)$}&\multicolumn{2}{c}{$(v=11)$}	\\
\hline\noalign{\smallskip}
0	&103\,830&574\,(5)		&104\,196&634\,3\,(5)		&104\,546&193\,(5)		&105\,158&038\,(10)	\\
1	&103\,836&646\,(5)		&104\,222&511\,2\,(5)		&104\,553&237\,(5)		&105\,167&538\,(5)       \\
2	&103\,848&722\,(5)		&104\,273&802\,1\,(5)		&104\,567&587\,(5)		&105\,186&262\,(5)       \\
3	&103\,866&724\,(5)		&104\,349&206\,9\,(5)		&104\,590&099\,(5)		&105\,213&063\,(5)       \\
4	&103\,890&579\,(5)		&104\,445&043\,9\,(5)		&104\,623&724\,(5)		&105\,247&353\,(5)        \\
5	&103\,920&228\,(5)		&104\,548&420\,1\,(10)		&104\,680&260\,(5)		&105\,288&197\,(5)       \\
6	&\multicolumn{2}{c}{}		&104\,548&463\,7\,(10)		&104\,787&933\,(5)		&105\,335&657\,(5)  	\\		
										
\hline\noalign{\smallskip}
$N$&\multicolumn{2}{c}{$(v=12)$}&\multicolumn{2}{c}{$(v=13)$}&\multicolumn{2}{c}{$(v=14)$}&\multicolumn{2}{c}{$(v=15)$}	\\
\hline\noalign{\smallskip}										        	
0	&105\,531&930\,(1)		&105\,977&638\,(2)		&106\,505&926\,(2)		&106\,985&727\,(2)		\\
1	&105\,548&839\,(1)		&105\,988&775\,(1)		&106\,517&373\,(2)		&106\,998&539\,(2)              \\
2	&105\,582&454\,(1)		&106\,011&360\,(1)		&106\,540&059\,(2)		&107\,024&061\,(2)              \\
3	&105\,632&067\,(1)		&106\,046&122\,(1)		&106\,573&643\,(2)		&107\,062&045\,(2)              \\
4	&105\,695&980\,(1)		&106\,094&346\,(1)		&106\,617&761\,(2)		&107\,112&023\,(2)              \\
5	&105\,771&154\,(1)		&106\,158&770\,(1)		&106\,672&213\,(2)		&\multicolumn{2}{c}{}              \\
6	&105\,853&413\,(1)		&\multicolumn{2}{c}{}		&\multicolumn{2}{c}{}		&\multicolumn{2}{c}{}       	\\
7	&105\,935&102\,(10)		&\multicolumn{2}{c}{}		&\multicolumn{2}{c}{}		&\multicolumn{2}{c}{}           \\

\hline\noalign{\smallskip}
$N$&\multicolumn{2}{c}{$(v=16)$}&\multicolumn{2}{c}{$(v=17)$}&\multicolumn{2}{c}{$(v=18)$}&\multicolumn{2}{c}{$(v=19)$}	\\
\hline\noalign{\smallskip}											
0	&107\,472&321\,(2)		&107\,980&549\,(4)		&108\,480&832\,(4)		&108\,972&673\,(6)	\\
1	&107\,484&157\,(2)		&107\,991&879\,(4)		&108\,492&373\,(4)		&108\,984&013\,(6)   	\\
2	&107\,507&890\,(2)		&108\,014&491\,(4)		&108\,515&378\,(4)		&109\,006&729\,(6)  	\\
3	&107\,543&643\,(2)		&108\,048&307\,(4)		&108\,549&680\,(4)		&109\,040&717\,(6)   	\\
4	&107\,591&577\,(2)		&108\,093&270\,(4)		&108\,595&044\,(4)		&109\,085&952\,(6)   	\\
5	&107\,651&815\,(2)		&108\,149&381\,(4)		&108\,651&153\,(4)		&109\,142&203\,(6)   	\\

\hline\noalign{\smallskip}
$N$&\multicolumn{2}{c}{$(v=20)$}&\multicolumn{2}{c}{$(v=21)$}&\multicolumn{2}{c}{}&\multicolumn{2}{c}{}	\\
\hline\noalign{\smallskip}											
0	&109\,466&309\,(6)		&109\,958&259\,(6)		&\multicolumn{2}{c}{}		&\multicolumn{2}{c}{}\\						
1	&109\,477&214\,(10)		&109\,968&971\,(6)		&\multicolumn{2}{c}{}		&\multicolumn{2}{c}{}\\					
2	&109\,498&981\,(6)		&109\,990&320\,(6)		&\multicolumn{2}{c}{}		&\multicolumn{2}{c}{}\\					
3	&109\,531&610\,(6)		&110\,022&206\,(6)		&\multicolumn{2}{c}{}		&\multicolumn{2}{c}{}\\					
4	&109\,575&054\,(6)		&110\,064&518\,(6)		&\multicolumn{2}{c}{}		&\multicolumn{2}{c}{}\\					
5	&109\,629&317\,(6)		&\multicolumn{2}{c}{}		&\multicolumn{2}{c}{}		&\multicolumn{2}{c}{}\\					

\end{longtable}
}

The FT-derived value for the \Dm\ \efstate\ ($v=0, N=2$) level energy is compared to the accurate measurement results from Doppler-free two-photon \EFX\ spectroscopy by Hannemann \etal~\cite{Hannemann2006}, Yiannopoulou \etal~\cite{Yiannopoulou2006}, and recently Niu \etal~\cite{Niu2014}. The comparison yields energy differences of $-0.000\,2\,(2)$ \wn, $0.000\,0\,(8)$ \wn, and $-0.000\,03\,(2)$ \wn, respectively.
For these comparisons, the ground state \xstate, $(v=0, N=2)$ and $(v=1, N=2)$ level energies used were taken from the theoretical values of Komasa \etal~\cite{Komasa2011} in addition to the \EFX\ Q(2) transition energies determined from the respective investigations in Refs.~\cite{Yiannopoulou2006,Hannemann2006,Dickenson2013}.
Note that the \efstate ($v=0, N=0,1$) level energies are not included in the comparison as these are the anchor levels used as absolute references to the \xstate\ ground state.
These comparisons for \Dm confirm the agreement in the case of \Hm \efstate\ ($v=0, N=2-5$) level energies as discussed in Refs.~\cite{Salumbides2008, Bailly2010} using identical methodologies described here.
In addition, a comparison for \Hm\ $EF (v=0, N = 6 -12)$ FT results with that of \EFX\ UV spectroscopy in Ref.~\cite{Salumbides2011} shows very good agreement. Similarly, a comparison with the \Hm\ results of Dickenson \etal~\cite{Dickenson2012}, detecting transitions up to the $EF(v=10)$ level shows very good agreement that is limited by the experimental accuracy of the latter study. The favorable comparisons for the case of \Hm give confidence in the uncertainty estimated for the present \Dm case.

\begin{figure}
\center
\resizebox{0.7\textwidth}{!}{\includegraphics{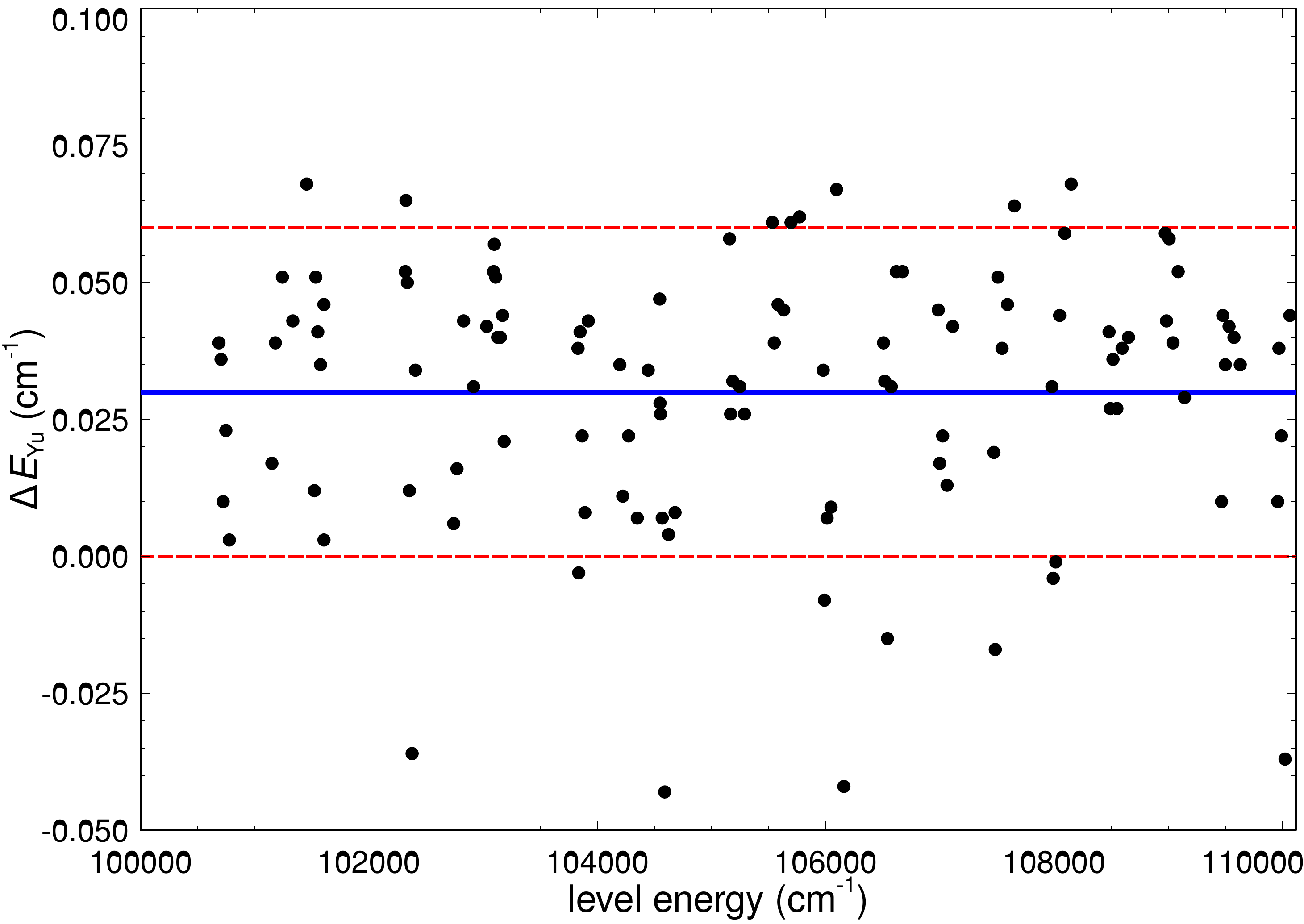}}
\caption{
Difference $\Delta E_\mathrm{Yu} = E_\mathrm{FT} - E_\mathrm{Yu}$  (in cm$^{-1}$) between the level energies between the present dataset and those from Ref.~\cite{Yu1994}. The solid line indicates the average while the dashed lines indicate the $\pm 1\sigma$ standard deviation of $\Delta E_\mathrm{Yu}$.
}
\label{ComparisonYuDressler}
\end{figure}

A comparison of the present level energies with the comprehensive data set from Yu and Dressler~\cite{Yu1994} was carried out for $EF, v>0$, with a graphical representation of the differences shown in Fig.~\ref{ComparisonYuDressler}.
For \Dm, the compilation of Yu and Dressler~\cite{Yu1994} was based on the experimental results compiled by Freund \etal~\cite{Freund1985} and those of Senn \etal~\cite{Senn1986}. Yu and Dressler had already applied a correction of -0.14 \wn\ to the term values in consideration of their the \abin\ investigations.
In Fig.~\ref{ComparisonYuDressler}, the present level energies $E_\mathrm{FT}$ are systematically higher by $\sim 0.03$ cm$^{-1}$ with respect to those of Ref.~\cite{Yu1994}. The standard deviation of 0.03 \wn\ for the difference $\Delta E_\mathrm{Yu}$ is consistent with the estimated statistical uncertainty of the experimental data used, i.e. 0.05 \wn in Ref.~\cite{Freund1985} and $\sim0.02$ \wn in Ref.~\cite{Senn1986}.

\begin{figure}
\center
\resizebox{0.7\textwidth}{!}{\includegraphics{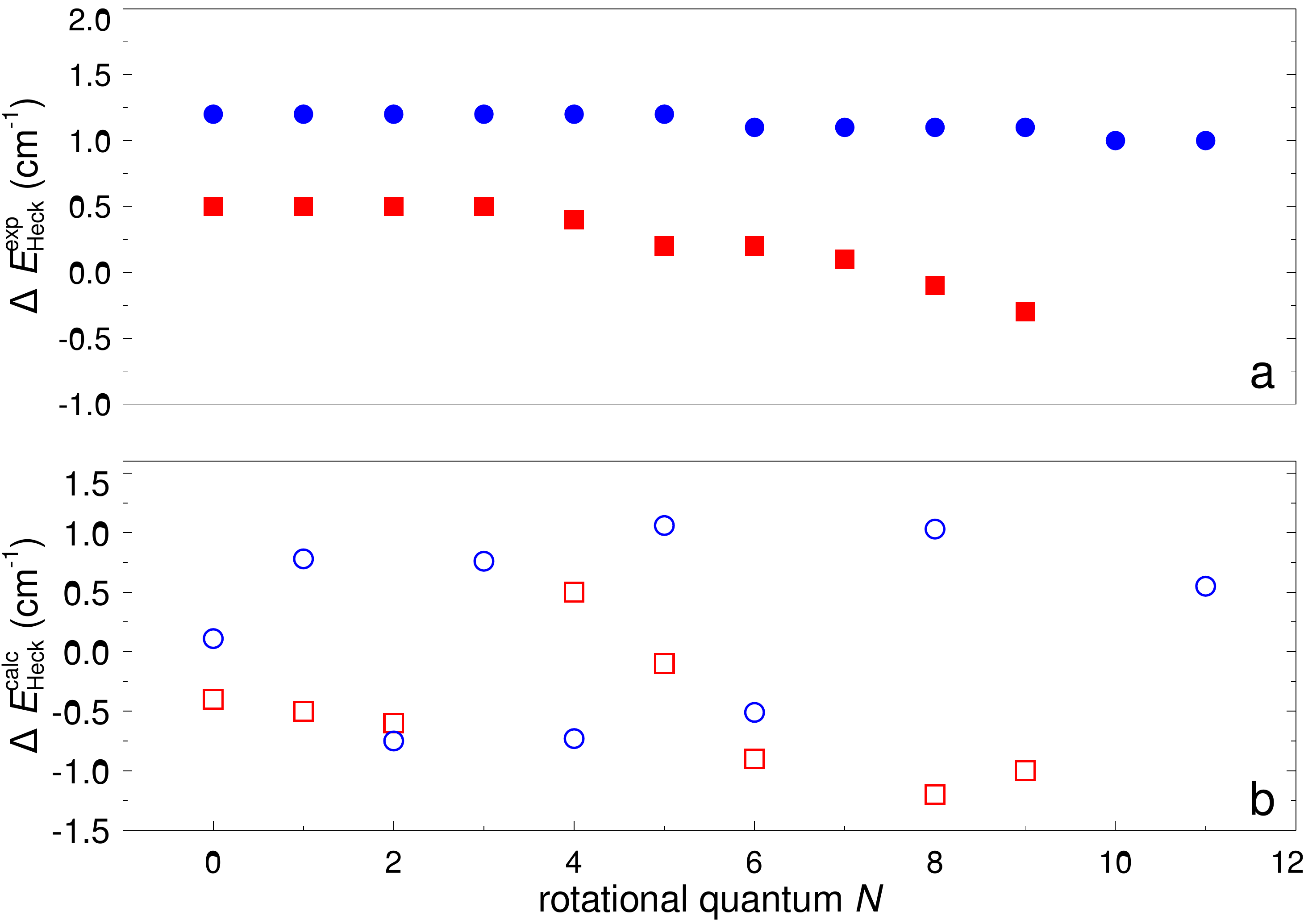}}
\caption{
Difference $\Delta E_\mathrm{Heck}^\mathrm{exp, calc} = E_\mathrm{FT} - E_\mathrm{Heck}^\mathrm{exp, calc}$ (in cm$^{-1}$) between the level energies from the present dataset and those from Ref.~\cite{Heck1995} plotted against the rotational quantum number $N=0-11$.
Datapoints for \efstate, $v=0$ are denoted by circles, while those for \efstate, $v=3$ ($E1$) by squares.
Comparison with experimental level energies $E_\mathrm{Heck}^\mathrm{exp}$ are shown in (a), while a comparison with their \abin\ calculations $E_\mathrm{Heck}^\mathrm{calc}$ is shown in (b).
}
\label{ComparisonHeck}
\end{figure}

Heck \etal~\cite{Heck1995} observed \Dm transitions between the \efstate $v'=0,1$ and \xstate $v''=0$ states involving high rotational quantum numbers up to $N=26$. The result of a comparison of the present results with that of Ref.~\cite{Heck1995} is plotted in Fig.~\ref{ComparisonHeck}. The upper panel (a) is a comparison with their experimental values, the lower panel (b) shows a comparison with their \emph{ab initio} calculations, for $EF, v=0, 1$ (represented by circles and squares, respectively) for levels with rotational quantum numbers up to $N=11$.
The observed energy differences are within the estimated experimental uncertainty by Heck \etal\ of better than 2 \wn~\cite{Heck1995}.
The \abin calculations performed by Heck \etal\ in the same study~\cite{Heck1995} are the most accurate calculations for \Dm \efstate level energies.
Although the accuracy was not specified for the \abin\ values, the comparison in Fig.~\ref{ComparisonHeck}(b) suggests that the calculations are accurate within 1 \wn in the energy range accessed.

{
\begin{table}\small
\caption{
Level energies (term values) in the \bstate\ state of D$_2$ for vibrational levels $v = 0-8$.
Values are in \wn\ with estimated ($1\sigma$) uncertainties indicated in between parentheses.
\label{B_levels}
}
\begin{center}
{\footnotesize
\begin{tabular}{c@{\hspace{20pt}}l@{\hspace{20pt}}l@{\hspace{20pt}}l}
\hline
\hline\noalign{\smallskip}
$N$&\multicolumn{1}{l}{$(v=0)$}&\multicolumn{1}{l}{$(v=1)$}&\multicolumn{1}{l}{$(v=2)$}\\
\hline\noalign{\smallskip}
0	&90\,633.471\,(1)       &91\,575.809\,(1)		&92\,498.689\,(1)	\\
1	&90\,653.192\,(1)       &91\,594.804\,(1)		&92\,517.052\,(1)	\\
2	&90\,692.535\,(1)       &91\,632.706\,(1)		&92\,553.698\,(1)	\\
3	&90\,751.307\,(1)       &91\,689.341\,(1)		&92\,608.464\,(1)	\\
4	&90\,829.218\,(1)       &91\,764.448\,(1)		&92\,681.114\,(1)	\\
5	&90\,925.892\,(1)       &91\,857.685\,(1)		&92\,771.336\,(1)	\\
6	&91\,040.868\,(1)       &91\,968.639\,(1)		&92\,878.745\,(1)	\\
7	&91\,173.614\,(1)       &92\,096.825\,(1)		&93\,002.899\,(1)    	\\
8	&91\,323.524\,(1)       &92\,241.695\,(1)    		&93\,143.285\,(1)	\\
9	&91\,489.940\,(1)       &92\,402.648\,(1)   		&93\,299.364\,(1)	\\
10	&91\,672.198\,(10)    	&92\,579.039\,(5)    		&	\\
\hline\noalign{\smallskip}
$N$&\multicolumn{1}{l}{$(v=3)$}&\multicolumn{1}{l}{$(v=4)$}&\multicolumn{1}{l}{$(v=5)$}\\
\hline\noalign{\smallskip}						
0	&93\,403.321\,(1)	&94\,290.405\,(1)	&95\,160.388\,(1)	\\
1	&93\,421.117\,(1)	&94\,307.683\,(1)	&95\,177.183\,(1)	\\
2	&93\,456.633\,(1)	&94\,342.167\,(1)	&95\,210.710\,(1)	\\
3	&93\,509.723\,(1)	&94\,393.720\,(1)	&95\,260.832\,(1)	\\
4	&93\,580.164\,(1)	&94\,462.133\,(1)	&95\,327.358\,(1)	\\
5	&93\,667.664\,(1)	&94\,547.135\,(1)	&95\,410.033\,(1)	\\
6	&93\,771.870\,(1)	&94\,648.390\,(1)	&95\,508.539\,(1)	\\
7	&93\,892.368\,(1)	&94\,765.515\,(1)	&95\,622.505\,(1)	\\
8	&94\,028.682\,(1)	&94\,898.060\,(1)	&95\,751.516\,(5)	\\
9	&94\,180.302\,(1)	&95\,045.539\,(2) 	&95\,895.107\,(5)	\\
\hline\noalign{\smallskip}
$N$&\multicolumn{1}{l}{$(v=6)$}&\multicolumn{1}{l}{$(v=7)$}&\multicolumn{1}{l}{$(v=8)$}	\\
\hline\noalign{\smallskip}
0	&96\,013.575\,(1)	&96\,850.205\,(1)	&97\,670.466\,(1)	\\
1	&96\,029.919\,(1)	&96\,866.118\,(1)	&97\,685.972\,(1)	\\
2	&96\,062.543\,(1)	&96\,897.890\,(1)	&97\,716.929\,(1)	\\
3	&96\,111.325\,(1)	&96\,945.398\,(1)	&97\,763.223\,(1)	\\
4	&96\,176.077\,(1)	&97\,008.468\,(1)	&97\,824.687\,(1)	\\
5	&96\,256.558\,(1)	&97\,086.868\,(1)	&97\,901.099\,(1)	\\
6	&96\,352.470\,(1)	&97\,180.312\,(1)	&97\,992.191\,(1)	\\
7	&96\,463.457\,(2)	&97\,288.469\,(2)	&98\,097.634\,(2)	\\
\hline\hline
\end{tabular}}
\end{center}
\end{table}}

\begin{figure}
\center
\resizebox{0.7\textwidth}{!}{\includegraphics{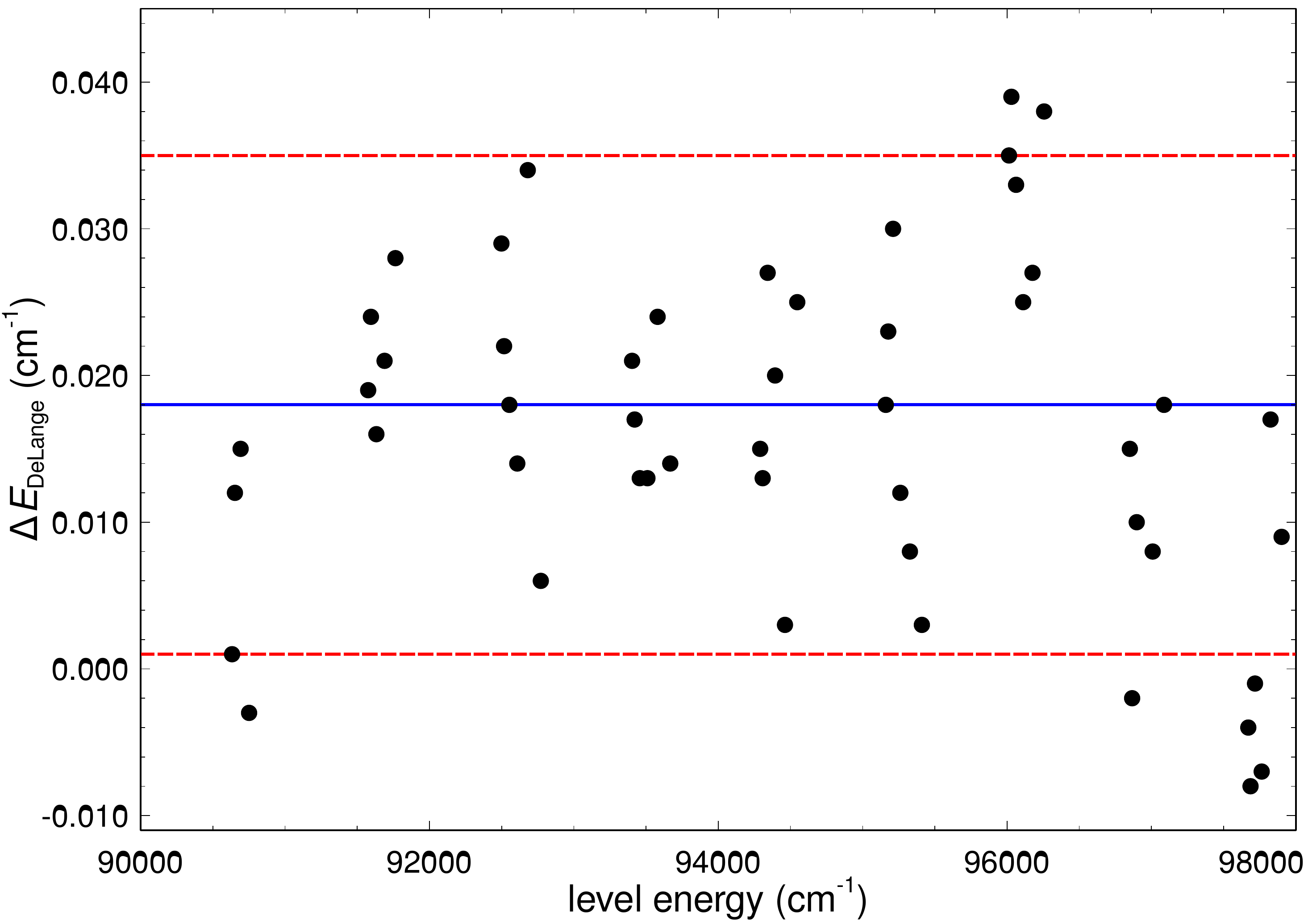}}
\caption{
Level energy differences $\Delta E_\mathrm{DeLange} = E_\mathrm{FT} - E_\mathrm{DeLange}$  (in \wn) of \bstate\ levels between the present dataset and those from Ref.~\cite{DeLange2012}. The solid line indicates the average while the dashed lines indicate the $\pm 1\sigma$ standard deviations of $\Delta E_\mathrm{DeLange}$.
}
\label{ComparisonDeLange}
\end{figure}

\section{\bstate\ level energies}

The \Dm \bstate level energies derived from the present FT data set are listed in Table~\ref{B_levels}, covering vibrational quantum numbers $v=0-8$, with rotational states as high as $N=10$.
The $B$ levels are referenced to the $EF$ anchor levels in the FT analysis, and are connected to \xstate, $(v=0, N=0)$ ground state in a straightforward manner.
Following the electronic symmetry of \bstate, ortho-\Dm levels have odd-$N$ while para-\Dm levels have even-$N$ quantum numbers.
Most level energies in Table~\ref{B_levels} are accurate to 0.001~\wn while the least accurate have uncertainties better than 0.01~\wn.

The recent work of de Lange \etal~\cite{DeLange2012} presents an accurate and comprehensive data set for \Dm \bstate\ level energies obtained from \BX\ XUV FT spectroscopy using a synchrotron source.
The database in Ref.~\cite{DeLange2012} included all bound level energies in the \bstate\ potential well, covering vibrational quantum numbers $v=0-51$ with rotational quantum numbers $N=0-5$.
An extensive comparison was carried out by de Lange \etal with the previous results by Hinnen \etal~\cite{Hinnen1994}, which showed that the values of the latter are systematically higher by 0.01 \wn. De Lange \etal also made a comparison with the semi-empirical study of Abgrall \etal~\cite{Abgrall1999}, based on the experimental results reported by Freund \etal~\cite{Freund1985}. The results of Abgrall \etal were shown to be shifted by 0.2 \wn higher with respect to the De Lange \etal values. This is not too surprising considering the estimated uncertainty of 0.5 \wn for the absolute calibration by Freund \etal~\cite{Freund1985}.
A direct comparison of the present results with the Abgrall \etal values~\cite{Abgrall1999} show the same trend as the comparison presented by de Lange \etal for the \bstate levels as expected.

The comparison of the present data set and Ref.~\cite{DeLange2012} reveals that the values of the latter are 0.02 \wn lower than the present data set (see Fig.~\ref{ComparisonDeLange}). This offset is still within the estimated uncertainties of 0.03 \wn by de Lange \etal~\cite{DeLange2012}, nevertheless, the systematic trend is notable. We attribute this to systematic effects in the absolute energy calibration of the synchrotron XUV FT data which relies on laser data for correction. In the case of Ref.~\cite{DeLange2012}, the XUV laser results from Roudjane \etal~\cite{Roudjane2008} for the \Dm \bstate, $v=9-11$ bands were utilized. In applying these corrections for the XUV FT, it is seen that systematic deviations increase the farther a certain transition is from the calibration line, and is probably limited by the FT relative energy calibration. In order to extract the most accurate results from the synchrotron XUV FT data, as is pursued for example in Ref.~\cite{Niu2013}, a regular coverage of calibration lines throughout the full spectrum is then required.

{
\begin{table}[t]\small
\caption{
\Dm\ Lyman (B $^1\Sigma^+_u$ -- X $^1\Sigma^+_g$) wavelengths.
R- and P-branch transitions to the \bstate, ($v = 0 - 8, N$) rovibrational levels from the lowest \xstate, ($v = 0$) vibrational level.
The wavelengths are given in nm with estimated uncertainties of $1\times10^{-6}$ nm.
The \xstate, ($v=0, N=0-5$) level energies are taken from Ref.~\cite{Komasa2011}.
\label{Lyman}
}
\begin{center}{\footnotesize
\begin{tabular}{c@{\hspace{18pt}}l@{\hspace{18pt}}l@{\hspace{30pt}}c@{\hspace{18pt}}l@{\hspace{18pt}}l}
\hline
\hline\noalign{\smallskip}
$N$	&\multicolumn{1}{c}{R($N$)}	&\multicolumn{1}{c}{P($N$)}	&$N$	&\multicolumn{1}{c}{R($N$)}	&\multicolumn{1}{c}{P($N$)}\\
\hline\noalign{\smallskip}
\multicolumn{3}{c}{$v=0$}						&\multicolumn{3}{c}{$v=1$}				     \\
\noalign{\smallskip}
0	&110.310\,511		&			&0	&109.176\,499		&		     \\
1	&110.335\,386		&110.407\,337		&1	&109.202\,583		&109.270\,476\\
2	&110.409\,106		&110.528\,839		&2	&109.277\,348		&109.390\,356\\
3	&110.531\,553		&110.698\,794		&3	&109.400\,652		&109.558\,555\\
4	&110.702\,525		&110.916\,894		&4	&109.572\,267		&109.774\,756\\
5	&110.921\,739		&111.182\,759		&5	&109.791\,872		&110.038\,561\\
\noalign{\smallskip}
\multicolumn{3}{c}{$v=2$}						&\multicolumn{3}{c}{$v=3$}				     	\\
\noalign{\smallskip}
0	&108.088\,182		&			&0	&107.042\,180		&			\\
1	&108.115\,218		&108.179\,555		&1	&107.069\,990		&107.131\,141   \\
2	&108.190\,687		&108.297\,793		&2	&107.145\,931		&107.247\,750   \\
3	&108.314\,433		&108.464\,125      	&3	&107.269\,839		&107.412\,172   \\
4	&108.486\,203		&108.678\,230      	&4	&107.441\,451		&107.624\,082   \\
5	&108.705\,660		&108.939\,701      	&5	&107.660\,409		&107.883\,071   \\
\noalign{\smallskip}
\multicolumn{3}{c}{$v=4$}						&\multicolumn{3}{c}{$v=5$}				     	\\
\noalign{\smallskip}
0	&106.035\,899		&			&0	&105.067\,199   	&			\\
1	&106.064\,350		&106.122\,612		&1	&105.096\,189		&105.151\,800	\\
2	&106.140\,603		&106.237\,619		&2	&105.172\,638		&105.265\,246  \\
3	&106.264\,484		&106.400\,124		&3	&105.296\,361		&105.425\,852  \\
4	&106.435\,722		&106.609\,804		&4	&105.467\,079		&105.633\,301  \\
5	&106.653\,953		&106.866\,243		&5	&105.684\,422		&105.887\,175  \\
\noalign{\smallskip}                                                                                                                    
\multicolumn{3}{c}{$v=6$}						&\multicolumn{3}{c}{$v=7$}				     	\\
\noalign{\smallskip}
0	&104.134\,213		&			&0	&103.235\,272		&			\\
1	&104.163\,670		&104.216\,827		&1	&103.265\,131		&103.316\,006	\\
2	&104.240\,224		&104.328\,755      	&2	&103.341\,730		&103.426\,466  \\
3	&104.363\,694		&104.487\,500      	&3	&103.464\,880		&103.583\,389  \\
4	&104.533\,796		&104.692\,738      	&4	&103.634\,297		&103.786\,460  \\
5	&104.750\,152		&104.944\,059      	&5	&103.849\,604		&104.035\,264  \\
\noalign{\smallskip}                                                                                                                    
\multicolumn{3}{c}{$v=8$}													     	\\
\noalign{\smallskip}
0	&102.368\,844		&			\\				
1	&102.399\,058		&102.447\,800	\\					
2	&102.475\,652		&102.556\,840  \\    					
3	&102.598\,436		&102.711\,993  \\    					
4	&102.767\,125		&102.912\,943  \\    					
5	&102.981\,334		&103.159\,282  \\    					
\hline\hline
\end{tabular}}
\end{center}
\end{table}
}

\section{Lyman band wavelengths}

Derived \Dm\ Lyman transition wavelengths connecting the \bstate ($v=0-8,N=0-5$) levels to the \xstate ($v=0,N$) levels, for the R- and P- branches, are listed in Table~\ref{Lyman} for $v=0-8, J=0-5$ as a line list for future application.
The transition wavelengths are derived using the present \bstate level energies and the ground state level energies from the calculations of Komasa \etal~\cite{Komasa2011}.
The wavelengths have relative accuracies of $1\times10^{-8}$ or better, limited by the accuracy of the \bstate level energies.
We note that although the Lyman values listed in Table~\ref{Lyman} are only for the \xstate ($v=0,N$) levels, transition energies connecting vibrationally-excited ground state levels can equally well be derived, e.g. using the values of Komasa \etal~\cite{Komasa2011}.

Gabriel \etal~\cite{Gabriel2009} determined \Hm, HD and \Dm Lyman transitions to high rovibrational levels in the \bstate state. A comparison show that the \Dm values of Gabriel \etal~\cite{Gabriel2009} are systematically higher by 0.23 \wn with respect to the present data set, with a scatter of 0.05 \wn. Note that this is at the accuracy limit estimated by Gabriel \etal~\cite{Gabriel2009} of 0.2 \wn.

\section{Conclusions}
\label{conclusions}
Accurate level energies for the \Dm\ \bstate, $v=0-8$ and \efstate\, $v=0-21$ bands are presented.
The present data set takes advantage of the extensive range covered by a highly-accurate FT spectroscopic investigation on \Dm excited state transitions, and the absolute accuracy \EFX\ UV spectroscopy to connect to the ground state.
For \bstate and \efstate levels connected with strong transitions, the accuracy is improved by more than order of magnitude with respect to previous studies.
The energies the \bstate levels obtained here will be useful calibration lines in XUV FT spectra, e.g. Ref.~\cite{DeLange2012}. 
The accurate \efstate level energies will be important in future molecular tests of QED in \Dm vibrationally- and rotationally-excited quantum levels as in Ref.~\cite{Salumbides2011}.

\section*{Acknowledgment}
We thank Arno de Lange and Evelyne Roueff for sending their respective results in digital form and useful discussions on the comparisons.

\section*{References}

\bibliographystyle{model1a-num-names}

\end{document}